\documentclass{article}
\usepackage{times}
\usepackage{amsfonts}
\usepackage[pdfmark,colorlinks]{hyperref}
\begin{document}
\title{Complex Moduli of Physical Quanta}
\author{Jos\'e M. Isidro\\
Instituto de F\'{\i}sica Corpuscular (CSIC--UVEG)\\
Apartado de Correos 22085, Valencia 46071, Spain\\
{\tt jmisidro@ific.uv.es}}

\maketitle

\begin{abstract}

Classical mechanics can be formulated using a symplectic structure on classical phase space, while quantum mechanics requires a complex--differentiable structure on that same space. Complex--differentiable structures on a given real manifold are often not unique. This letter is devoted to analysing the dependence of the notion of a quantum on the complex--differentiable structure chosen on classical phase space.

\end{abstract}

\tableofcontents

\section{Introduction}\label{labastidahijoputa}

The name {\it M--theory}\/ is used to refer to an 11--dimensional quantum theory of gravity which reduces to $N=1$ supergravity at low energy (for a review see, {\it e.g.}, ref. \cite{VAFA}). It is a feature of this theory that it unifies a number of apparently unrelated models (of superstrings and supergravities) as different limits of one single entity. Often, these limits can be mapped into other such limits by means of transformations called {\it dualities}. The latter are best understood as (apparently) different descriptions of one and the same underlying physics. It has been argued \cite{VAFA} that understanding M--theory dualities requires a reassessment of the concept of {\it classical}\/ vs. {\it quantum}\/ already at the level of a finite number of degrees of freedom, {\it i.e.}, before moving on to fields, strings and branes. Moreover, the BFSS matrix model \cite{BFSS} is claimed to provide a fundamental, quantum--mechanical description of M--theory (for a review see, {\it e.g.}, ref. \cite{NICOLAI}). Motivated by these considerations we have proposed some approaches \cite{GOLM, QMGCS} to the problem of implementing duality transformations within the context of a finite number of degrees of freeedom. Admittedly, this is a finite truncation of the infinite number of degrees of freedom present in M--theory, but it is nevertheless worth studying. Then one can approach the dynamics by studying classical phase space ${\cal C}$, and then moving on to quantisation by (more or less) unambiguous procedures. In what follows it is pointed out that one can exploit the (often unnoticed) ambiguities underlying the quantisation procedure in order to implement dualities.

\section{Deformations of a complex--differentiable structure}\label{labastidamecagoentuputasombra}

{}For the sake of simplicity we will limit our analysis to $\mathbb{R}^{2n}$ as a classical phase space. This will suffice to illustrate our conclusions without loss of generality. The space $\mathbb{R}^{2n}$ has the Darboux coordinates $q^j$, $p_j$, $j=1,\ldots, n$, and the symplectic form
\begin{equation}
\omega=\sum_{j=1}^n{\rm d}q^j\wedge{\rm d}p_j.
\label{hijoputaluisibanez}
\end{equation}
The corresponding quantum operators satisfy 
\begin{equation}
[Q^j,P_l]={\rm i}\delta^{j}_{l}.
\label{kabronluisibanez}
\end{equation}
We endow $\mathbb{R}^n$ with the Euclidean metric 
\begin{equation}
g=\frac{1}{2}\sum_{j=1}^n\left(({\rm d}q^j)^2+({\rm d}p_j)^2\right).
\label{putoluisibanez}
\end{equation}
Complex $n$--dimensional space $\mathbb{C}^n$ has the holomorphic coordinates
\begin{equation}
z^j=\frac{1}{\sqrt{2}}\left(q^j+{\rm i}p_j\right), \qquad j=1,\ldots, n,
\label{labastidaquetepartaunrayo}
\end{equation}
and is endowed with the same metric as $\mathbb{R}^{2n}$, now Hermitian instead of real bilinear, 
\begin{equation}
g=\sum_{j=1}^n {\rm d}\bar z^j{\rm d}z^j=\vert\vert {\rm d}z\vert\vert^2.
\label{mekagoentuputakaraluisibanez}
\end{equation}
The corresponding quantum operators 
\begin{equation}
A^j=\frac{1}{\sqrt{2}}\left(Q^j+{\rm i}P_j\right),\qquad
(A^j)^+=\frac{1}{\sqrt{2}}\left(Q^j-{\rm i}P_j\right)
\label{luisibanezquetepartaunrayo}
\end{equation}
satisfy
\begin{equation}
[A^j,(A^l)^+]=\delta^{jl}.
\label{luisibanezmekagoentuputakaramarikondemierda}
\end{equation}

The space $\mathbb{C}^n$ has a moduli space of complex structures that are compatible with a given orientation. This moduli space is denoted ${\cal M}(\mathbb{C}^n)$; it is the symmetric space
\begin{equation}
{\cal M}(\mathbb{C}^n)=SO(2n)/U(n).
\label{amico}
\end{equation}
This is a compact space of real dimension $n(n-1)$. Here the embedding of $U(n)$ into $SO(2n)$ is given by
\begin{equation}
A+{\rm i}B\longrightarrow\left(\begin{array}{cc}
A&B\\
-B&A
\end{array}\right),
\label{labastidaquetelametanporculo}
\end{equation}
where $A+{\rm i}B\in U(n)$ with $A, B$ real, $n\times n$ matrices \cite{HELGASON}. 

Let us see how the symmetric space (\ref{amico}) appears as a moduli space of nonequivalent complex structures. Consider the Euclidean metric $g$ of eqn. (\ref{putoluisibanez}). Requiring rotations to preserve the orientation, the isometry group of $g$ is $SO(2n)$. In the complex coordinates of eqn. (\ref{labastidaquetepartaunrayo}), $g$ becomes the Hermitian form (\ref{mekagoentuputakaraluisibanez}), whose isometry group is $U(n)$. Notice that we no longer impose the condition of unit determinant, since $U(n)=SU(n)\times U(1)$ and $g$ is invariant under the $U(1)$ action $z^j\rightarrow {\rm e}^{{\rm i}\alpha}z^j$, $j=1,\ldots, n$, for all $\alpha\in\mathbb{R}$. Now every choice of orthogonal axes $x^j$, $y^j$ in $\mathbb{R}^{2n}$, {\it i.e.}, every element of $SO(2n)$, defines a complex structure on $\mathbb{R}^{2n}$ upon setting 
\begin{equation}
w^j=\frac{1}{\sqrt{2}}\left(x^j+{\rm i}y^j\right),\qquad j=1,\ldots, n. 
\label{kakaxramallo}
\end{equation}
Generically the $w^j$ are related nonbiholomorphically with the $z^j$, because the orthogonal transformation 
\begin{eqnarray}
z^j\longrightarrow w^j&=&\sum_{m=1}^n\left(R^j_mz^m+S^j_{m}\bar z^{m}\right)\nonumber\\
\bar z^{j}\longrightarrow \bar w^{j}&=&\sum_{m=1}^n\left(\bar R^{j}_{ m}\bar z^{m}+\bar S^{j}_{m} z^{m}\right),
\label{mierdaxramallo}
\end{eqnarray}
while satisfying the orthogonality conditions
\begin{equation}
\sum_{j=1}^n\left(R^j_m\,\bar R^{j}_{l}+S^j_{l}\, \bar S^{j}_m\right)=\delta_{ml},\qquad
\sum_{j=1}^nR^j_m\,\bar S^{j}_l=0=\sum_{j=1}^nS^j_{ m}\,\bar R^{j}_{l},
\label{kakaxbarbon}
\end{equation}
need not satisfy the Cauchy--Riemann conditions 
\begin{equation}
\frac{\partial \bar w^{j}}{\partial z^m}=\bar S^{j}_m=0=S^j_{m}=\frac{\partial  w^{j}}{\partial \bar z^{m}}.
\label{mierdaxcesargomez}
\end{equation}
However, when eqn. (\ref{mierdaxcesargomez}) holds, the transformation (\ref{mierdaxramallo}) is not just orthogonal but also unitary. 
Therefore one must divide $SO(2n)$ by the action of the unitary group $U(n)$, in order to obtain the parameter space for rotations that truly correspond to inequivalent complex structures on $\mathbb{R}^{2n}\simeq \mathbb{C}^n$. Nonbiholomorphic complex structures on $\mathbb{C}^n$ are 1--to--1 with rotations of $\mathbb{R}^{2n}$ that are {\it not}\/ unitary transformations.

When $n=1$ the moduli space (\ref{amico}) reduces to a point. Therefore on the complex plane $\mathbb{C}$ there exists a unique complex structure, that we can identify as the one whose holomorphic atlas consists of the open set $\mathbb{C}$ endowed with the holomorphic coordinate $z=(q+{\rm i}p)/\sqrt{2}$. Physically this corresponds to the 1--dimensional harmonic oscillator. Consider now $n$ independent harmonic oscillators, where ${\cal C}=\mathbb{C}^n=\mathbb{C}\times{(n)\atop\cdots}\times\mathbb{C}$. Although it is never explicitly stated, the complex structure on this product space is always understood to be the $n$--fold Cartesian product of the unique complex structure on $\mathbb{C}$. Obviously, removing the requirement of compatibility between the complex structure and the orientation chosen, we duplicate the number of complex structures. See ref. \cite{BRUZZO} for a detailed treatment of the theory of deformations of complex structures.

In ref. \cite{QMGCS} we have argued that, when classical phase space ${\cal C}$ is a generalised complex manifold, we have a natural setup for duality transformations that complements the one presented in ref. \cite{GOLM}. For background material on generalised complex structures see ref. \cite{GUALTIERI}. Roughly speaking, any generalised complex manifold ${\cal C}$ splits {\it locally}\/ as the product of a complex manifold times a symplectic manifold. A more precise statement is as follows. A point $x\in{\cal C}$ is said {\it regular}\/ if the Poisson structure $\omega^{-1}$ has constant rank in a neighbourhood of $x$. Then any regular point in a generalised complex manifold ${\cal C}$ has a neighbourhood that is equivalent to the product of an open set in $\mathbb{C}^k$ with an open set in $\mathbb{R}^{2n-2k}$, the latter endowed with its standard symplectic form. The nonnegative integer $k$ is called the {\it type}\/ of the generalised complex structure. The limiting case $k=0$ corresponds to ${\cal C}$ being a symplectic manifold, while $k=n$ corresponds 
to ${\cal C}$ being a complex manifold.

A generalised complex manifold ${\cal C}$ of real dimension $2n$ has a moduli space of generalised complex structures \cite{GUALTIERI}
\begin{equation}
{\cal M}({\cal C})=SO(2n,2n)/U(n,n).
\label{labastidakabronazo}
\end{equation}
In the context of generalised complex structures, the appearance of the noncompact groups $SO(2n,2n)$ and $U(n,n)$ instead of their compact partners $SO(2n)$ and $U(n)$ is quite natural. Mathematically \cite{GUALTIERI} it is motivated in the presence of a metric of indefinite signature $(+\ldots^{(2n)}\ldots+,-\ldots^{(2n)}\ldots-)$. Physically \cite{QMGCS} this metric gives rise to the {\it Planck cone}. Within $\mathbb{R}^{2n}\oplus\mathbb{R}^{2n}$, the latter is defined by 
\begin{equation}
\sum_{j=1}^n\left((Q^j)^2+(P_j)^2\right)-\sum_{j=1}^n\left((q^j)^2+(p_j)^2\right)=0.
\label{luiibanezmekagoentuputakalva}
\end{equation}
A duality transformation has been identified in ref. \cite{QMGCS} as {\it a crossing of the Planck cone}. This approach to duality transformations has the added bonus that it bears a strong resemblance with the theory of relativity. In fact a duality is nothing but the relativity of the notion of a quantum. Many textbooks on special relativity take the constancy of the speed of light as their starting point. Mathematically this can be recast as the invariance of the light--cone under Lorentz transformations. The light--cone separates physical particles from tachyons, the cone itself corresponding to massless particles. The Lorentz group $SO(1,3)$ arises naturally in this setup. In our context we have the group $SO(2n,2n)$ instead, with the Planck cone above replacing the light--cone.

\section{Discussion}\label{kakaparaluisibanez}

Complex--differentiable structures on classical phase spaces ${\cal C}$ have a twofold meaning. Geometrically they define  
complex differentiability, or analyticity, of functions on complex manifolds such as ${\cal C}$. Quantum--mechanically they  
define the notion of a quantum, {\it i.e.}, an elementary excitation of the vacuum state \cite{PERELOMOV}. In this letter we have elaborated on this latter meaning. The mathematical possibility of having two or more nonbiholomorphic complex--differentiable structures on a given classical phase space leads to the physical notion of a quantum--mechanical duality, {\it i.e.}, to the relativity of the notion of an elementary quantum. This relativity is understood as the dependence of a quantum on the choice of a complex--differentiable structure on ${\cal C}$. One can summarise this fact in the statement that a quantum is a complex--differentiable structure on classical phase space \cite{GOLM}. A duality arises as the possibility of having two or more, apparently different, descriptions of the same physics. These facts imply that the concept of a quantum is not absolute, but relative to the quantum theory used to measure it \cite{VAFA}. In particular {\it classical}\/ and {\it quantum}\/, for long known to be intimately related \cite{MATONE}, are not necessarily always the same for all observers on phase space. 

In this letter we have analysed the dependence of the notion of a quantum on the complex--differentiable structure chosen on classical phase space. When the latter is $\mathbb{R}^{2n}$ we have established the existence a moduli space of nonbiholomorphic complex structures ${\cal M}(\mathbb{C}^n)=SO(2n)/U(n)$. Moving around within ${\cal M}(\mathbb{C}^n)$ we obtain nonequivalent definitions of complex differentiability. The transformations between observers carrying nonbiholomorphic complex structures are nonholomorphic. Hence the corresponding observers do not agree on the notion of an elementary quantum. In the particular case of $\mathbb{C}^n$, variations in the complex structure are 1--to--1 with variations in the symplectic structure. The viewpoint that one can consider nonequivalent symplectic structures on classical phase space has been exploited in ref. \cite{MERCED}. A related problem, dealing with the existence of a continuum of nonequivalent Hamiltonian structures, has ben analysed in ref. \cite{MEX}.

In the light of recent developments in the theory of complex manifolds \cite{GUALTIERI} it is possible to interpolate between the symplectic category and the complex category. In other words, one can interpolate between classical and quantum mechanics. Correspondingly we have  extended our analysis of the dependence of physical quanta on complex moduli to the case when classical phase space is a generalised complex manifold. In this latter case a duality can be given a pictorial interpretation as a crossing of the Planck cone. The latter separates symplectic from complex, {\it i.e.}, classical from quantum. The suggestion of implementing duality transformations within the classical and quantum mechanics of a finite number of degrees of freedom was put forward in the light of developments in M--theory \cite{VAFA}. Regardless of strings and branes, however, an interesting picture emerges for dualities once one realises the following point. Whatever one's favourite choice is for a quantum theory of gravity (for a recent review see, {\it e.g.}, ref. \cite{ASHTEKAR}), quantising gravity is dual to relativising the notion of a quantum. In rendering the concept of a quantum observer--dependent we are, in a sense, quantising gravity. Thus, rather than looking around for the quanta of gravity, we are searching for the moduli of quanta.

{\bf Acknowledgements}

It is a great pleasure to thank J. de Azc\'arraga for encouragement and support. The author thanks Max-Planck-Institut f\"ur Gravitationsphysik (Potsdam, Germany) where this work was begun, for hospitality. This work has been partially supported by research grant BFM2002--03681 from Ministerio de Ciencia y Tecnolog\'{\i}a, by research grant GV2004-B-226 from Generalitat Valenciana, by EU FEDER funds, by Fundaci\'on Marina Bueno and by Deutsche Forschungsgemeinschaft.

\end{document}